\documentclass[final]{ws-procs9x6}

\newcommand{\ftn}{\footnotesize}

\newcommand{\hepph}[1]{{\tt hep-ph/#1}}

\newcommand{\astroph}[1]{{\tt astro-ph/#1}}

\newcommand{\grqc}[1]{{\tt gr-qc/#1}}

\newcommand\astp[3]{{\sl Astropart.\ Phys.\ }{\bf #1}, #3 (#2)}

\newcommand\jhep[3]
        {{\sl J. High Energy Phys.\ }{\bf #1}, #3 (#2)}

\newcommand\jcap[3]
        {{\sl J. Cosmol. Astropart. Phys.\ }{\bf #1}, #3 (#2)}

\newcommand\npb[3]
        {{\sl Nucl.\ Phys.\ }{\bf B#1}, #3 (#2)}

\newcommand\plb[3]
        {{\sl Phys.\ Lett.\ }{B \bf #1}, #3 (#2)}

\newcommand\prd[3]
        {{\sl Phys.\ Rev.\ }{D \bf #1}, #3 (#2)}

\newcommand\prl[3]
        {{\sl Phys.\ Rev.\ Lett.\ }{\bf #1}, #3 (#2)}

\newcommand{\vrho}{{\mbox{$\bar\rho$}}}

\newcommand{\sv}{{\mbox{$\langle \sigma v \rangle$}}}

\newcommand{\brhofi}{{\mbox{$\bar\rho_{\phi_{\rm I}}$}}}
\newcommand{\brhoqi}{{\mbox{$\bar\rho_{q_{\rm I}}$}}}

\newcommand{\brhof}{{\mbox{$\bar\rho_{\phi}$}}}
\newcommand{\brhoq}{{\mbox{$\bar\rho_{q}$}}}

\newcommand{\brhoRi}{{\mbox{$\vrho_{_{\rm RI}}$}}}

\def\openep{\leavevmode\hbox{\normalsize$\iota$\kern-3.8pt$^$-}}
\def\vtau{\leavevmode\hbox{\normalsize$\tau$\kern-5.pt$\iota$}}
\def\vtauf{\leavevmode\hbox{\ftn$\tau$\kern-4.pt$\iota$}}
\def\btau{\leavevmode\hbox{\normalsize$\tilde\tau$\kern-5.pt$\iota$}}
\def\btauf{\leavevmode\hbox{\ftn$\tilde\tau$\kern-4.pt$\iota$}}

\def\beq{\begin{equation}}
\def\eeq{\end{equation}}
\def\bea{\begin{eqnarray}}
\def\eea{\end{eqnarray}}
\def\openep{\leavevmode\hbox{\normalsize$\epsilon$\kern-4.pt$\epsilon$}}

\begin{document}

\title{CDM Abundance in non-Standard Cosmologies}

\author{C. Pallis}

\address{School of Physics and Astronomy,
The University of Manchester,\\
Manchester M13 9PL, UNITED KINGDOM \\
E-mail: Constantinos.Pallis@manchester.ac.uk}

\maketitle

\pagestyle{headings}

\thispagestyle{plain}\markboth{C. Pallis}{CDM Abundance in
non-Standard Cosmologies}\setcounter{page}{1}

\abstracts{The relic density of a cold dark matter (CDM) candidate
is calculated in the context of three non-standard cosmological
scenaria and its value is compared with the one obtained in the
standard regime. In particular, we consider the decoupling of the
CDM particle during: (i) A decaying-particle dominated phase or
(ii) a kinetic-energy dominated phase or (iii) the decay of a
massive particle under the complete or partial domination of
kination. We use plausible values (from the viewpoint of
supersymmetric models) for the mass and the thermal averaged cross
section times the velocity of the cold relic and we investigate
scenaria of equilibrium and non-equilibrium production. In the
case (i) a low reheat temperature, in the range $(1-20)~{\rm
GeV}$, significantly facilitates the achievement of an acceptable
CDM abundance. On the other hand, the presence of kination in the
case (ii) can lead to an enhancement of the CDM abundance up to
three orders of magnitude. The latter enhancement can be avoided,
in the case (iii). In a such situation, the temperature turns out
to be frozen to a plateau value which is, mostly, lower than about
$40~{\rm GeV}$.}

\section{Introduction} \label{sec:intro}
This review is based on Refs. [1,2,3]. We recall the calculation
of the CDM abundance (Sec.~\ref{omega}) in the context of the
Standard scenario (SC) (Sec.~\ref{sc}) and we show how this
calculation is modified in three different cases: When we consider
a Low Reheating Scenario (LRS)\cite{scnr} (Sec.~\ref{lrs}), a
Quintessential Scenario (QKS)\cite{jcapa} (Sec.~\ref{qks}) or a
Kination-Dominated Reheating (KRS)\cite{qui} (Sec.~\ref{krs}). A
comparison between the various scenarios is displayed in
Table~\ref{table} whereas in Table ~\ref{table1} we present some
representative combinations of the parameters which produce the
central observational value of the CDM abundance. Throughout the
subscript or superscript $0$ [I] is referred to present-day values
(except for the coefficient $V_0$) [to the onset of each scenario]
and $\bar \rho_{i}=\rho_{i}/\rho^0_{\rm c}$ where $\rho^0_{\rm
c}=8.099\times10^{-47}h^2~{\rm GeV^4}$ with\cite{wmap} $h=0.73$.

\begin{table}[!b]
\tbl{Standard vs non-Standard Scenaria} {\footnotesize
\begin{tabular}{@{}cccc@{}}
\hline {} &{} &{} &{}\\[-1.5ex] {\bf SC}& {\bf LRS}& {\bf QKS} &
{\bf KRS} \\ \hline {} &{} &{} &{} \\[-1.5ex]
$\brhoq=\brhof=0$&$\brhofi\gg\brhoRi,~\brhoq=0$&
$\brhoqi\gg\brhoRi,~\brhof=0$&$\brhoqi\gg\brhofi\gg\brhoRi$\\
$H\propto T^2$& $H\propto T^4$&$H\propto T^3$&$H\propto T^3$\\
$T\propto R^{-1}$&$T\propto R^{-3/8}$&$T\propto R^{-1}$&$T={\rm
cst}$\\
$sR^3={\rm cst}$&$sR^3\neq{\rm cst}$&$sR^3={\rm
cst}$&$sR^3\neq{\rm cst}$\\
$N_\chi=0$&$N_\chi\neq0$&$N_\chi=0$&$N_\chi\neq0$
\\ [1ex]\hline
\end{tabular} \label{table} }
\vspace*{-10pt}
\end{table}

\section{The CDM abundance} \label{omega}
In light of the recent WMAP3 results\cite{wmap}, the relic density
of any CDM candidate $\chi$, $\Omega_{\chi}h^2$, is to satisfy the
following range of values:
\beq \label{cdmb}\Omega_{\chi}h^2
=0.1045_{-0.0095}^{+0.0075}~~\Rightarrow~~0.08\lesssim
\Omega_{\chi}h^2\lesssim0.12~~\mbox{at $95\%$ c.l.}\eeq
We concentrate our presentation mainly on the Lightest SUSY
Particle (LSP) which is stable within the SUSY models with
$R$-parity conservation\cite{goldberg} and consists the most
popular, promising and natural CDM candidate\cite{candidates}.

The calculation of $\Omega_{\chi}h^2$ is based on the formula:
\beq\Omega_{\chi}h^2 = 2.741 \times 10^8\ Y_0\
m_{\chi}/\mbox{GeV}, ~~{\rm where}~~Y_0=n_{\chi}^0/s_0,\eeq
$s\propto T^3$ is the entropy density, $m_{\chi}$ is the mass of
$\chi$ and $n_{\chi}$ is the number density of $\chi$'s, which
satisfies a Boltzmann Equation (BE), provided that ${\chi}$'s
achieve {\it kinetic} equilibrium with plasma. The form of the BE
depends on the scenario under consideration. In general, it
depends on: {\sf (i)} The Hubble parameter, $H=\sqrt{\rho_{_{\rm
BG}}}/\sqrt{3}m_{_{\rm P}}$ ($m_{_{\rm P}}=M_{\rm P}/\sqrt{8\pi}$,
with $M_{\rm P}$, the Planck scale) with $\rho_{_{\rm BG}}$, the
background energy density, {\sf (ii)} the equilibrium number
density of $\chi$'s, $n_{\chi}^{\rm eq}$, which obeys the
Maxwell-Boltzmann statistics:
\beq n_{\chi}^{\rm eq}(x)=\frac{g}{(2\pi)^{3/2}}
m_{\chi}^3\>x^{3/2}\>e^{-1/x}P_2(1/x),\quad\mbox{where}\quad
x=T/m_{\chi}<1,\label{neqx} \eeq
with $g=2$ the number of degrees of freedom of ${\chi}$ and
$P_n(z)=1+(4n^2-1)/8z$, {\sf (iii)} the thermal-averaged cross
section times velocity of $\chi$'s, $\langle \sigma v \rangle$,
which can be mostly expanded as: $\langle \sigma v \rangle=a+bx$.
We focus on the case $\langle \sigma v \rangle=a$ with
$10^{-15}~{\rm GeV}^{-2}\leq\langle\sigma v\rangle\leq10^{-7}~{\rm
GeV}^{-2}$ which can be naturally produced within SUSY
models\cite{scnr,jcapa,qui}.

Moreover, two fundamental cases of $\chi$-production can be
singled out\cite{tkolb}: $\chi$'s do or do not maintain {\it
chemical} equilibrium with plasma. In the first case (EP) the
current value of $n_{\chi}/s$ follows $n^{\rm eq}_{\chi}/s$ and at
some $T=T_{\rm F}$, $n_{\chi}/s$ becomes larger than $n^{\rm
eq}_{\chi}/s$. On the other hand, in the case of non-EP,
$n_{\chi}/s\gg n^{\rm eq}_{\chi}/s$ (non-EPI) or $n_{\chi}/s\ll
n^{\rm eq}_{\chi}/s$ (non-EPII) at least at the point of the
maximal $\chi$ production\cite{qui,tkolb}.

\section{The Standard Cosmological Scenario (SC)} \label{sc}

According to the SC\cite{goldberg}, $\chi$'s {\sf (i)} are
produced through thermal scatterings, {\sf (ii)} reach {\it
chemical} equilibrium with plasma and {\sf (iii)} decouple from
the cosmic fluid at a temperature $T=T_{\rm F}\sim 15~{\rm GeV}$
during the radiation-dominated (RD) era. The consequences of the
assumptions above are: {\sf (i)} The form of the relevant BE is
(dot denotes derivative w.r.t the cosmic time),
\beq \dot n_{\chi}+3Hn_{\chi}+\langle \sigma v \rangle \left(
n_{\chi}^2 - n_{\chi}^{\rm eq2}\right)=0\label{BEs} \eeq
which can be solved numerically (or semi-analytically using the
freeze-out procedure\cite{scnr,jcapa,qui}) with initial condition
$n_{\chi}(x=1)=n^{\rm eq}_{\chi}(x=1)$ or $n_{\chi}(x=1)=0$, {\sf
(ii)} the required \sv\ is $\langle\sigma
v\rangle\gtrsim10^{-20}~{\rm GeV}^{-2}$ (note that with
$\langle\sigma v\rangle\simeq2.9\times10^{-29}~{\rm GeV}^{-2}$ and
$n_{\chi}(x=1)=0$, we can obtain $\Omega_{\chi}h^2=0.1$ if we
allow for non-EPII), {\sf (iii)} the cosmological evolution during
the $\chi$ decoupling is RD and so, $\rho_{_{\rm
BG}}\simeq\rho_{_{\rm R}}\propto T^4$. Therefore, $H\propto T^2$
and $T\propto R^{-1}$, where $R$ is the scale factor of the
universe (see Table~\ref{table}).

In this context, the $\Omega_{\chi}h^2$ calculation depends only
on two parameters: $m_\chi~\mbox{and}~\langle \sigma v \rangle$.
As shown in Table~\ref{table1} (1st column), $\langle \sigma v
\rangle\sim 10^{-9}~{\rm GeV}^{-2}$ can ensure acceptable
$\Omega_{\chi}h^2$'s. Such a requirement strongly restricts the
parameter space of many particle models. However, this picture can
drastically change if one or more assumptions of the SC are
lifted.


\begin{sidewaystable}
\tbl{Combinations of parameters leading to $\Omega_{\chi}h^2=0.1$
in the Standard and non-Standard Scenaria.
\label{table1}}{\footnotesize
{\begin{tabular}{@{}cccccccccccccc@{}} \hline
\multicolumn{14}{c}{}\\[-1.5ex] {}&{} &\multicolumn{1}{c}{\bf SC}
&{}&\multicolumn{3}{c}{\bf LRS}&{} &\multicolumn{2}{c}{\bf QKS}&{}
&\multicolumn{3}{c}{\bf KRS}\\[0.25ex] \hline
\multicolumn{14}{c}{}\\[-1.5ex]
$\chi$-Pro- &\multicolumn{13}{c}{}\\
duction:&{}
&EP&{}&EP&non-EPII&EP&{}&EP&EP&{}&non-EPI&non-EPII&EP\\ [0.25ex]
\hline \multicolumn{14}{c}{}\\[-1.5ex]
$\sv~(\rm GeV^{-2})$ &&$2\times 10^{-9}$
&&$10^{-10}$&$10^{-10}$&$10^{-8}$&&$3\times10^{-6}$&$3.6\times10^{-7}$&&
$10^{-10}$&$10^{-10}$&$10^{-10}$\\ [0.25ex] \hline
\multicolumn{14}{c}{}\\[-1.5ex]
$T_\phi~(\rm GeV)$
&&$-$&&$5.5$&$0.001$&$5.5$&&$-$&$-$&&$30$&$30$&$5.5$\\[0.25ex]
$N_\chi$
&&$-$&&$0$&$10^{-3}$&$5\times10^{-5}$&&$0$&$0$&&$10^{-6}$&$0$&$0$\\
[0.25ex] \hline \multicolumn{14}{c}{}\\[-1.5ex]
$\Omega_q^{\rm NS}$
&&$-$&&$-$&$-$&$-$&&$0.1$&$0.001$&&$0.02$&$3\times10^{-9}$&$2.4\times10^{-15}$
\\ [0.25ex] \hline \multicolumn{14}{c}{}\\[-1.5ex]
$\log\brhofi$&&$-$&&$95.62$&$95.62$&$95.62$&&$-$&$-$&&$69.3$&$73.815$&$76.5$
\\[0.25ex]
$T_{\rm PL}~(\rm GeV)$
&&$-$&&$-$&$-$&$-$&&$-$&$-$&&$1.2$&$16$&$32.5$
\\[0.25ex] \hline \multicolumn{14}{c}{}\\ [-1.5ex]
$\Omega_{\chi}h^2|_{\rm
SC}$&&$0.1$&&$1.9$&$1.9$&$0.023$&&$9\times10^{-5}$&$7\times10^{-4}$&&$1.9$&$1.9$&$1.9$
\\ [1ex] \hline
\end{tabular}}}
\begin{tabnote}
We fix $m_{\chi}=350~{\rm GeV}$ for the results presented in this
table. We also use $m_\phi=10^6~{\rm GeV}$ in the LRS and KRS and
$\lambda=0.5$ in the QKS and KRS (note, however, that the results
on $\Omega_\chi h^2$ are $\lambda$-independent). Recall that in
the LRS, $T_\phi=T_{\rm RH}$ and the results on $\Omega_\chi h^2$
are $\brhofi$-independent and remain invariant for fixed
$N_{\chi}m^{-1}_\phi$ (and, obviously, $T_\phi,~m_\chi,~\sv$) --
however, the type of the $\chi$-production does depend separately
on $N_{\chi}$ and $m_\phi$. In each case, shown is the type of the
$\chi$-production and the obtained value of $\Omega_\chi h^2$ in
the SC, $\Omega_\chi h^2|_{\rm SC}$.
\end{tabnote}
\end{sidewaystable}

\section{The Low Reheating Scenario (LRS)} \label{lrs}

The modern cosmo-particles theories are abundant in scalar massive
particles (e.g. moduli, PQ-flatons, dilatons) which can decay when
$H$ becomes equal to their mass creating episodes of reheating. In
the LRS, we assume that such a scalar particle $\phi$, with mass
$m_\phi$, decays with a rate $\Gamma_\phi$ into radiation,
producing an average number $N_{\chi}$ of $\chi$'s, rapidly
thermalized. The energy densities of $\phi$ and the produced
radiation and $\rho_{_{\rm R}}$ and $\rho_\phi$ and
$n_{\tilde\chi}$, satisfy the following BEs
($\Delta_\phi=(m_\phi-N_{\tilde\chi}m_{\tilde\chi})/m_\phi$):
\bea && \dot \rho_\phi+3H\rho_\phi+\Gamma_\phi
\rho_\phi=0,\label{rf}\\
&& \dot \rho_{_{\rm R}}+4H\rho_{_{\rm R}}-\Gamma_\phi \Delta_\phi
\rho_\phi-2m_{\chi}\langle \sigma v \rangle \left( n_{\chi}^2 -
n_{\chi}^{\rm eq2}\right)=0, \label{rR}\\
&& \dot n_{\chi}+3Hn_{\chi}+\langle \sigma v \rangle \left(
n_{\chi}^2 - n_{\chi}^{\rm eq2}\right)-\Gamma_\phi N_{\chi}
n_\phi=0,\vspace*{-4.5mm}\label{nx}\eea
The system above can be solved, imposing the following initial
conditions:\vspace*{-5.mm}
\begin{equation}
H_{\rm
I}=m_\phi~\Rightarrow~\brhofi=m_\phi^2/H_0^2~~\mbox{and}~~\brhoRi=\bar\rho_{\chi_{\rm
I}}=0. \vspace*{-1.5mm}\label{init}
\end{equation}
Our investigation verifies that the reheating process is not
instantaneous\cite{turner}. Until its completion -- i.e.,
$\rho_\phi(T_{\rm RH})=\rho_{_{\rm R}}(T_{\rm RH})$ --, the
maximal temperature, $T_{\rm max}=f(\Gamma_\phi,\brhofi)$, can
become much larger than the so-called reheat
temperature\cite{riotto}, $T_{\rm RH}$. Also, for $T>T_{\rm RH}$,
$\rho_{_{\rm BG}}\simeq\rho_{\phi}~\Rightarrow~ H\propto T^4$ with
$T \propto R^{-3/8}$ and an entropy production occurs (see
Table~\ref{table}).

The free parameters of the LRS are:
$m_{\chi},~\langle \sigma v
\rangle,~\Gamma_{\phi},~m_\phi,~N_{\chi},~\brhofi.$
However, the results on $\Omega_{\chi}h^2$ do not depend on the
explicit value of $\brhofi$ as long as $T_{\rm RH}<T_{\rm
F}<T_{\rm max}$, and are invariant\cite{scnr,gondolon} for fixed
$N_{\chi}m^{-1}_\phi$ (and $T_\phi,~m_\chi,~\sv$). Moreover,
$\Gamma_{\phi}$ can be replaced by $T_\phi\simeq T_{\rm RH}$
through the relation\cite{qui}:\vspace*{-5.mm}
\begin{equation}
\Gamma_\phi =5\sqrt{\frac{\pi^3 g_{\rho*}(T_\phi)}{45}}
\frac{T_\phi^2}{M_{\rm P}}=\sqrt{\frac{5\pi^3
g_{\rho*}(T_\phi)}{72}} \frac{T_\phi^2}{m_{_{\rm P}}}
\cdot\label{GTrh}
\end{equation}
%
As shown in Table~\ref{table1}, we can obtain acceptable
$\Omega_\chi h^2$'s {\sf (i) } for relatively low \sv\ 's (2nd,
3rd columns) with low $T_{\rm RH}=5.5~{\rm GeV}$ and
$N_\chi=0$\cite{scnr,riotto} (EP) or much lower $T_{\rm RH}=1~{\rm
MeV}$ and $N_\chi\neq0$ (non-EPII) and {\sf (ii)} for larger \sv\
's (4th column) with larger $N_\chi$'s (EP).

\section{The Quintessential Scenario (QKS)} \label{qks}

Another role that a scalar field could play when it does not
couple to matter (contrary to $\phi$) is this of quintessence.
Such a field, $q$, satisfies the equation: \vspace*{-5.mm}
\beq \ddot q+3H\dot q+dV/dq=0,~~\mbox{where}~~V=V_0 e^{-\lambda
q/m_{_{\rm P}}}\label{qeq} \eeq
is the adopted potential\cite{wet} and undergoes three phases
during its cosmological evolution\cite{jcapa}: {\sf (i)} The
kination\cite{kination} dominated (KD) phase, where the energy
density of $q$ is $\rho_q=\dot q^2/2+V\simeq \dot q^2/2$, {\sf
(ii)} the frozen-field dominated (FD) phase, where $\rho_q$ is
constant and {\sf (iii)} the late-time attractor dominated (AD)
phase, with $w^{\rm fp}_q=\lambda^2/3-1$ for $\lambda<\sqrt{3}$.
Today we obtain a transition from the FD to the AD
phase\cite{jcapa,brazil}. Although this does not provide a
satisfactory resolution of the coincidence problem, the
observational data\cite{wmap}: \vspace*{-5.mm}
\beq \Omega^0_q=\Omega_{\rm
DE}=0.74\pm0.12~~\mbox{and}~~w_q<-0.83.\label{de} \eeq
can be reproduced with $\lambda\leq1.1$ and by conveniently
adjusting $V_0$\cite{jcapa,brazil}.

For a reasonable region of initial conditions ($q_{\rm I}=0$ and
$\Omega_q^{\rm I}=1$\cite{jcapa}), $\rho_q\simeq \dot q^2/2$ can
dominate over radiation, creating a totally KD era in conjunction
with the satisfaction of the nucleosynthesis (NS)
constraint\cite{oliven}, $\Omega_q^{\rm NS}\leq0.21$. As a
consequence, during the KD era we obtain: $\rho_{_{\rm
BG}}\simeq\rho_{q}~\Rightarrow~H\propto T^3$ with $T\propto
R^{-1}$ (see Table~\ref{table}). Combining eqs.~(\ref{BEs}) and
(\ref{qeq}) we can show that if the $\chi$-decoupling occurs
during the KD era, $\Omega_\chi h^2$ increases\cite{salati} w.r.t
$\Omega_\chi h^2|_{\rm SC}$ (see 5th and 6th column in
Table~\ref{table1}). This enhancement of $\Omega_\chi h^2$ turns
out to be a single-valued function of the quintessential parameter
at the eve of NS\cite{jcapa}, $\Omega_q^{\rm NS}$, for fixed
$m_\chi$ and $\langle \sigma v \rangle$. Therefore, the
$\Omega_{\chi}h^2$ calculation depends only on the parameters:
$m_{\chi},~\langle \sigma v \rangle,\ \Omega^{\rm
NS}_q~~(\lambda\leq1.1).$
%

\begin{figure}[!ht]
\centering
\includegraphics[width=4.25cm,angle=-90]{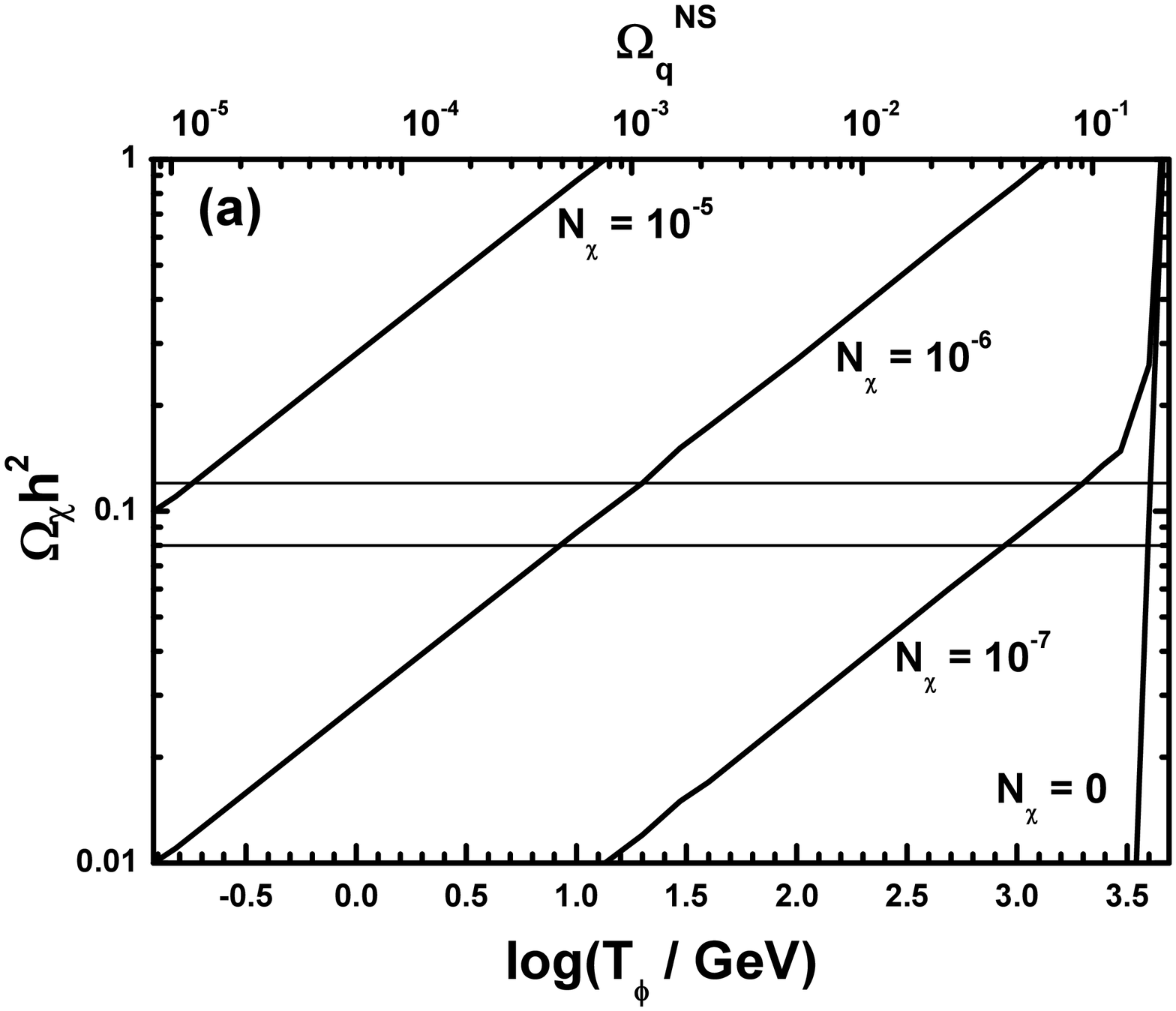}
\hspace*{-1.1cm}
\includegraphics[width=4.25cm,angle=-90]{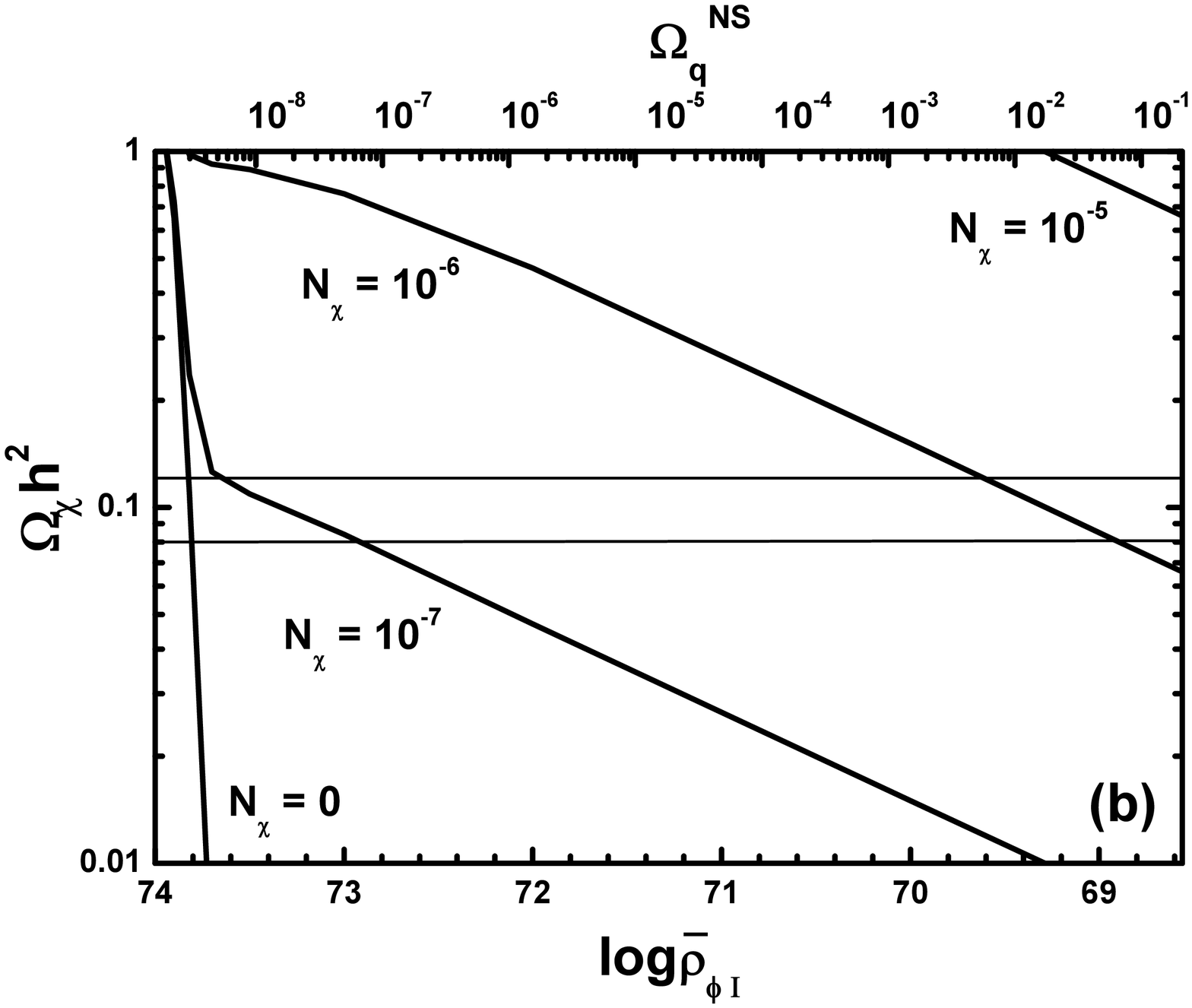}
\caption{\ftn $\Omega_{\tilde\chi}h^2$ versus $\log T_\phi$ (or
$\Omega_q^{\rm NS}$) for $\log\brhofi=70$ ~{\sf (a)} and
$\log\brhofi$ (or $\Omega_q^{\rm NS}$) for $T_\phi=30~{\rm
GeV}$~{\sf (b)}. We take $m_\phi=10^6~{\rm
GeV},~m_{\tilde\chi}=350~{\rm GeV},~\langle\sigma
v\rangle=10^{-10}~{\rm GeV^{-2}}$ and various $N_\chi$'s,
indicated on the curves. The CDM bounds of Eq.~(\ref{cdmb}) are,
also, depicted by the two thin lines.} \label{qom}
\end{figure}

\section{The Kination-Dominated Reheating (KRS)} \label{krs}

In view of the two previous situations, the obvious question would
be: What happens if we have both quintessence and low reheating?
Or, a low $T_{\rm RH}$ could assist us to the reduction of
$\Omega_{\chi}h^2$, in the presence of a KD phase? This
novel\cite{liddle} cosmological set-up can be analyzed by solving
the system of Eqs.~(\ref{rf})-(\ref{nx}) and (\ref{qeq}) with
constraint $\Omega^{\rm I}_q=1$ and initial conditions:
\begin{equation}
q_{\rm I}=0,~~H_{\rm I}=m_\phi~\Rightarrow~\dot q_{\rm
I}=\sqrt{2\rho^0_{\rm
c}}(m_\phi/H_0)~~\mbox{and}~~\brhoRi=\bar\rho_{\chi_{\rm I}}=0.
\label{init1}
\end{equation}
We can distinguish two types of $q$-domination, depending whether
$\phi$ decays before or after it becomes the dominant component of
the universe. In both cases, $\rho_{_{\rm
BG}}\simeq\rho_{q}~\Rightarrow~H\propto T^3$, entropy production
occurs and a prominent period of constant maximal temperature,
$T_{\rm PL}=f(\brhofi,\Gamma_\phi/m_\phi)$, arises\cite{qui} (see
Table~\ref{table}). The free parameters of this scenario are:
$ \lambda,\ \brhofi,\ m_\phi,\ T_\phi,\ N_{\chi},\ m_{\chi},\
\sv.$
As in the QKS the $\Omega_{\chi}h^2$ calculation is
$\lambda$-independent for $\lambda\leq1.1$ but unlike the LRS it
severely depends on $\brhofi$, and $T_\phi$ does not coincide with
the maximal temperature of the RD era.

The crucial difference between the KRS and the QKS is that
$\Omega_{\chi}h^2$ does not exclusively increase with $\Omega^{\rm
NS}_q$. Indeed, when an increase of $\Omega^{\rm NS}_q$ is
generated by an increase of $T_\phi$ (which results to an increase
of $T_{\rm PL}$), $\Omega_{\chi}h^2$ increases with $\Omega^{\rm
NS}_q$ -- see Fig.~\ref{qom}-{\sf (a)}. On the contrary, when the
increase of $\Omega^{\rm NS}_q$ is due to the decrease of
$\brhofi$ (which results to a decrease of $T_{\rm PL}$),
$\Omega_{\chi}h^2$ decreases as $\Omega^{\rm NS}_q$ increases --
see Fig.~\ref{qom}-{\sf (b)}. This is, because $n_\chi$ decreases
rapidly with $T_{\rm PL}$ due to the exponential suppression of
$n_\chi^{\rm eq}$ in Eq.~(\ref{neqx}).


As shown in Table~\ref{table1} (7th and 8th columns),
$\Omega_{\chi}h^2$ reaches the range of Eq.~(\ref{cdmb}) with
$N_\chi\sim(10^{-7}-10^{-5})$ when $T_{\rm PL}\ll T_{\rm F}$
(non-EPI) and with $N_\chi\sim0$ when $T_{\rm PL}\sim T_{\rm F}$
(non-EPII). As $\Omega^{\rm NS}_q$ decreases, $(T_{\rm PL}- T_{\rm
F})$ increases and $\Omega_{\chi}h^2$ approaches its value in the
LRS (9th column in Table~\ref{table1}).


\section{Conclusions} \label{clu}

We considered three deviations from the SC and we
showed\cite{scnr,jcapa,qui} that: {\sf (i)} In the LRS with
$T_{\rm RH}<20~{\rm GeV}$, $\Omega_{\chi}h^2$ decreases w.r.t its
value in the SC for low $N_\chi$'s and increases for larger
$N_\chi$'s. Both EP and non-EP are possible for commonly
obtainable $\sv$'s, {\sf (ii)} in the QKS, $\Omega_\chi h^2$
increases drastically (almost 3 orders of magnitude for
$\Omega_q^{\rm NS}$ close to its upper bound) {\sf (iii)} in the
KRS, $\Omega_{\chi}h^2$ becomes cosmologically interesting for
$N_\chi\sim(10^{-7}-10^{-5})$ when $T_{\rm PL}\ll T_{\rm F}$
(non-EPI), and for $N_\chi\sim0$ when $T_{\rm PL}\sim T_{\rm F}$
(non-EPII); EP is activated for $T_{\rm PL}>T_{\rm F}$ and the
results on $\Omega_{\chi}h^2$ approach their values in the LRS or
SC as $T_{\rm PL}$ increases well beyond $T_{\rm F}$.

\section*{Acknowledgments}
The author would like to thank the organizers of IDM2006 for their
invitation and the PPARC research grand PP/D000157/1 for financial
support.


\begin{thebibliography}{99}{\ftn

\bibitem{scnr} C. Pallis, \astp{21}{2004}{689} [\hepph{0402033}].

\bibitem{jcapa} C. Pallis, \jcap{10}{2005}{015} [\hepph{0503080}].

\bibitem{qui} C. Pallis, \npb{751}{2006}{129} [\hepph{0510234}].
%

\bibitem{wmap} D.N. Spergel {\it et al.}, \astroph{0603449}.

\bibitem{goldberg} H. Goldberg, \prl{50}{1983}{1419}; \\
J.R. Ellis {\it et al.}, \npb{238}{1984}{453}.

\bibitem{candidates} For other candidates see, e.g.,
G. Lazarides, \hepph{0601016}.


\bibitem{tkolb} E.W. Kolb, \hepph{9910311}.

\bibitem{turner} R.J. Scherrer and M.S. Turner, \prd{31}{1985}{681}.

\bibitem{riotto} J.~McDonald, \prd{43}{1991}{1063};\\ G.F. Giudice {\it et al.},
\prd{64}{2001}{023508} [\hepph{0005123}];\\ N. Fornengo {\it et
al.}, \prd{67}{2003}{023514} [\hepph{0208072}].

\bibitem{gondolon} G.~Gelmini and P.~Gondolo,
\prd{74}{2006}{023510} [\hepph{0602230}].

\bibitem{wet} C. Wetterich, \npb{302}{1988}{668}.

\bibitem{kination} B.~Spokoiny, \plb{315}{1993}{40}
[\grqc{9306008}];\\ M. Joyce, \prd{55}{1997}{1875}
[\hepph{9606223}].

\bibitem{brazil} U. Fran\c{c}a {\it et al.}, \jhep{10}{2002}{015}
[\astroph{0206194}];\\ C.L. Gardner, \npb{707}{2005}{278}
[\astroph{0407604}].


\bibitem{salati} P. Salati, \plb{571}{2003}{121}
[\astroph{0207396}].

\bibitem{oliven} R.H. Cyburt {\it et al.}, \astp{23}{2005}{313}
[\astroph{0408033}].

\bibitem{liddle} See, also, A. Liddle {\it et al.}, \prd{68}{2003}{043517}
[\astroph{0302054}];\\ B. Feng and M. Li, \plb{564}{2003}{169}
[\hepph{0212233}].




}


\end{thebibliography}
\end{document}